\documentclass[12pt]{article}

\usepackage{graphicx}

\title{\textbf{Is it a power law distribution? The case of economic contractions}}

\author{Salvador Pueyo\thanks{E-mail: spueyo@ic3.cat}\\\textit{\small{Institut Catal\`{a} de Ci\`{e}ncies del Clima (IC3), C/ Doctor Trueta 203, 08005 Barcelona,}}\\\textit{\small{Catalonia, Spain}}}

\date{}

\begin{document}

\maketitle

\begin{abstract}
\noindent One of the first steps to understand and forecast economic downturns is identifying their frequency distribution, but it remains uncertain. This problem is common in phenomena displaying power-law-like distributions. Power laws play a central role in complex systems theory; therefore, the current limitations in the identification of this distribution in empirical data are a major obstacle to pursue the insights that the complexity approach offers in many fields. This paper addresses this issue by introducing a reliable methodology with a solid theoretical foundation, the \textit{Taylor Series-Based Power Law Range Identification Method}. When applied to time series from 39 countries, this method reveals a well-defined power law in the relative per capita GDP contractions that span from 5.53\% to 50\%, comprising 263 events. However, this observation does not suffice to attribute recessions to some specific mechanism, such as self-organized criticality. The paper highlights a set of points requiring more study so as to discriminate among models compatible with the power law, as needed to develop sound tools for the management of recessions.\\

\noindent\textbf{Keywords:} \textit{Econophysics; economic crises; economic depressions; Pareto distribution; Bayesian hypothesis testing; self-organized criticality.}
\end{abstract}

\section{Introduction}

Equilibrium is a basic assumption of mainstream economic models. Such models are not the optimum tool when the political priority is to manage an intrinsically non-equilibrium phenomenon as is a recession. As a result, an increasing attention is paid to non-equilibrium models related to complex systems theory, especially those belonging to the new field of \textit{econophysics} \cite{Mantegna2000,Chakraborti1,Chakraborti2}. 

One of the properties to which complex systems research pays most attention is scale invariance, often in the form of some key variable $x$ displaying a power law distribution
\begin{equation}
\label{powerlaw}
f(x) = ( \frac{\tau-1}{x_{min}^{-\tau+1}-x_{max}^{-\tau+1}} ) x^{-\tau},
\end{equation}
where $f$ is the probability density function (PDF), the exponent $\tau$ is a constant and $x_{min}$, $x_{max}$ are the lower and upper bounds to the distribution (for fundamental reasons, power laws only apply over some limited range; Section \ref{basic}). Among the variables displaying a power law or power-law-like distributions, there are many different types of catastrophic events \cite{Pruessner}, such as earthquakes, landslides, storms, forest fires and some epidemics; there is a rich literature on mechanistic models attempting to explain this fact \cite{Pruessner}.

Despite theoretical suggestions that similar models would apply to economic downturns \cite{Bak_econ,Scheinkman,Krugman,Thurner,Xi}, and evidence of power laws in other economic variables \cite{Gabaix_powerlaws}, the attempts to test whether the sizes of GDP contractions actually follow a power law have been surprisingly rare, and they did not reach a definitive answer. The first step, carried out by Ormerod and Mounfield \cite{Ormerod2001}, was an exploratory analysis with inconclusive results, and the related analyses in refs.\ \cite{Ormerod2004,Xi} do not either give clearcut results (discussed in Section \ref{discusmet}). 

The difficulties in reaching strong conclusions are general in the literature dealing with this type of distribution in every field. The whole research program on power laws has been recently called into question \cite{Stumpf2012}. This is a serious challenge for complex systems theory and for econophysics: a recent assessment of what econophysics has contributed to economics \cite{Buchanan_econ} stated that \textit{More than anything, physicists have helped to establish empirical facts about financial markets} and went on to emphasize some presumed instances of self-similarity and power laws. However, there have also been some responses to this challenge. Also in recent times, there have been notable contributions for a sounder treatment of candidate power laws \cite{Clauset,DelucaCorral2013}, which might ultimately lead to a well-founded standard methodology for data analysis. The present contribution gives some further steps forward in this direction.

The first objective of this paper is to present a reliable method for power law data treatment with firm roots in probability theory: the \textit{TAylor SEries-based POwer LAw Range identification method} (TASEPOLAR). The second is to use TASEPOLAR to elucidate if relative GDP contractions display a power law distribution. Having obtained a positive result, the paper ends with a discussion on the possible causes and implications.

Similarly to refs.\ \cite{Ormerod2001,Ormerod2004,Xi}, the present paper considers contraction events during their whole duration. This differs from refs.\ \cite{Lee,Canning}, which dealt with year-to-year variations and found a distribution distinct from the power law. Barro and Jin \cite{BarroJin} and Barro and Urs\'{u}a \cite{BarroUrsua} (whose data I use) also studied complete events. However, they did not investigate the possibility of a power law in contraction size $x$, but in $y = x/(1-x)$. This amounts to hypothesizing a completely different distribution for $x$, which, as shown in Section \ref{rescomp}, does not fit the data as well as the power law.

\section{Theory: The Taylor Series-Based Power Law Range Identification Method (TASEPOLAR)}

\subsection{Basic features}
\label{basic}

There are at least two problems that make statistical inference especially challenging when dealing with a power law. First, that its properties are quite different from those of the Gaussian distribution, which is at the core of textbook statistics. Second, the fact that, for fundamental reasons \cite{Pueyo2011}, power laws apply within finite bounds $x_{min}$, $x_{max}$ and, often, these do not coincide with the smallest and largest possible value in the sample, i.e.\  Eq.\ (\ref{powerlaw}) might only apply to one part of the frequency distribution. For example, this is to be expected in our case. If Eq.\ (\ref{powerlaw}) applied to all drops in GDP $\geq 0$, there would be an infinite probability density for a relative contraction size $x \rightarrow 0$ (unless $\tau < 1$), i.e.\  there would be no fluctuations. Therefore, the PDF should flatten for small values. On the other hand, we cannot have $x_{max} \rightarrow \infty$ because a country cannot lose more than 100\% of its GDP. It is sometimes possible to use $x_{max} \rightarrow \infty$ as an approximation when $\tau > 2$, but this condition is not fulfilled according to our estimates. For $\tau \leq 2$, $x_{max} \rightarrow \infty$ would imply an infinite expectation \cite{Pueyo2011}. Therefore, the PDF necessarily bends downward for large values.

Let us take a sample of $N$ data $\{x\}$ and sort them in descending order, $x_{i} \geq x_{i+1}$ for $i$ from 1 to $N-1$. Let us call $D_{ij}$ to the subset of data such that $x_{i} \geq x \geq x_{j}$. TASEPOLAR tests the power law for different subsets $D_{ij}$ in order to choose an optimum range $[x_{min},x_{max}]$. With variations, this is also the schema followed by some other methods (Section \ref{discusmet}).

The main singularity of TASEPOLAR, which gives it its name, is the way to choose the hypothesis to be compared with the power law. Continuous and differentiable functions can be decomposed into a Taylor series. If we take a small enough range, any such function can be fitted by a straight line. Therefore, for most frequency distributions, if we take $\ln(x)$ vs $\ln(f(x))$ it will be trivially possible to fit a power law for a small enough range:
\[
\ln(f(x)) \approx \ln(a)-\tau\ln(x).
\]
If the range is slightly larger, we might need to add a second term to the Taylor series:
\begin{equation}
\label{Taylor}
\ln(f(x)) \approx \ln(a)-\tau\ln(x)-\psi[\ln(x)]^2.
\end{equation}
This function is equivalent to a truncated lognormal \cite{Pueyo2006}, with $\mu = -(\tau-1)/2\psi $ and $\sigma=1/\sqrt{2\psi}$. TASEPOLAR selects a range in which there is no evidence of a second term in the Taylor series expansion. Therefore, it performs a set of comparisons between the fit to a truncated power law and to a truncated lognormal.

The two distributions are compared by means of the likelihood ratio:
\begin{equation}
\label{lambda}
\lambda_{ij} = \frac{f(D_{ij}|H_{p},\hat{\tau}_{ij},x_{max}=x_{i},x_{min}=x_{j})}{f(D_{ij}|H_{l},\hat{\tau}_{ij},\hat{\psi}_{ij},x_{max}=x_{i},x_{min}=x_{j})},
\end{equation}
where $H_p$ is the hypothesis of a truncated power law, $H_l$ is the hypothesis of a truncated lognormal, and $\hat{\tau}_{ij}$, $\hat{\psi}_{ij}$ are the maximum likelihood estimates of $\tau$, $\psi$ obtained from $D_{ij}$ assuming $x_{max}=x_i$ and $x_{min}=x_j$. The interpretation of the likelihood ratio is discussed in Section \ref{interpret}.

TASEPOLAR selects a range $[i,j]$ that is as broad as possible while satisfying $\lambda_{ij}>\lambda_m$ for some given $\lambda_m$. This can be done automatically. However, in order to examine carefully whether the data are well-behaved, I used a semiautomatic, supervised method involving two stepwise sequences (details in Section \ref{aplic}). 

For this study, I chose $\lambda_m=0.99$. Note that, since the only difference between the two hypotheses is that the truncated lognormal allows us to tune one extra parameter when maximizing likelihood (i.e.\  they are nested hypotheses), $\lambda \leq 1$. For any large set of data that actually follow a power law, the probability of finding a value of $\lambda$ between 0.99 and 1 is $ \approx 0.11$ \cite{Wilks}. Therefore, when increasing (or decreasing) $i$ (or $j$) sequentially, $\lambda_{ij}$ follows a random walk and visits the region $\lambda_{ij}>\lambda_m$ with a frequency close to 0.11 if the power law hypothesis is correct (i.e.\  if $\psi = 0$). If $\psi \not= 0$, a trend will appear that will drive $\lambda$ far from this region.

Besides choosing a range that is well fitted by a power law, we want to make sure that this region is not just a slice of a truncated (or untruncated) lognormal too small for the influence of $\psi$ to be detected. We will do this by taking a larger region and testing whether it consists of a truncated lognormal. 

This involves two choices. First, the size of the larger region. In this study, all values smaller than $x_{min}$ were considered, in addition to the range from $x_{min}$ to $x_{max}$ (see Section \ref{test} on the reasons to exclude values larger than $x_{max}$ also when testing the lognormal). Second, since the lognormal hypothesis is tested against the hypothesis that the data display a power law in the range from $x_{min}$ to $x_{max}$ and some given distribution in the range $x<x_{min}$, the later distribution has to be specified. In this study, a power law was hypothesized also for $x<x_{min}$, with an exponent that can differ from the exponent in the other part of the PDF. 

Therefore, we are comparing two hypotheses for the set of data $x<x_{max}$. One of them ($H_d$) is that these data follow a \textit{double power law}, which consists of a power law with some exponent $\tau$ between $x_{min}$ and $x_{max}$, and a power law with some exponent $\tau '$ between $x_{min}'$ and $x_{max}'$, where $x_{min}'$ is the smallest value in the sample and $x_{max}'=x_{min}$, with no discontinuity (i.e.\  $f(x_{max}')$ as calculated in the lower range has to equal $f(x_{min})$ as calculated in the upper range). Given $x_{min}'$ and $x_{max}$, the double power law has three parameters, $\tau$, $\tau '$ and $x_{min}$, whose maximum likelihood estimators are sought (therefore, the value of $x_{min}$ used for this test does not have to coincide exactly with the one that we have previously fitted, and will generally be less conservative). The other hypothesis ($H_l$) is that the data follow a truncated lognormal in the range from $x_{min}'$ to $x_{max}$. Given these two limits, this distribution has two parameters ($\tau$ and $\psi$ in Eq.\ (\ref{Taylor}), or, equivalently, $\mu$ and $\sigma$) to be fitted by maximum likelihood estimation. The test uses the following likelihood ratio:
\begin{equation}
\label{lambda'}
\lambda_{ij} = \frac{f(D_{ij} | H_l,\hat{\tau}_{ij},\hat{\psi}_{ij},x_{max}=x_i,x_{min}'=x_j)}{f(D_{ij} | H_d,\hat{\tau}_{ij},\hat{\tau}_{ij}',\hat{x}_{min},x_{max}=x_i,x_{min}'=x_j)}.
\end{equation}
While a single truncated power law and a truncated lognormal are nested distributions, a double power law and a truncated lognormal are not. Therefore, in spite of the double power law having more parameters, the likelihood ratio will generally favour the truncated lognormal if the later is the distribution giving a correct description of the data. Moreover, the double power law is penalized in our test because of its extra parameter (Section \ref{interpret}). 

\subsection{Interpretation of the likelihood ratio}
\label{interpret}

TASEPOLAR uses likelihood ratios as a tool to compare hypotheses. More generally than in Eqs.\ (\ref{lambda}, \ref{lambda'}), the likelihood ratio can be expressed as
\begin{equation}
\label{lambdagen}
\lambda = \frac{f(D|H_1,\hat{p}_1)}{f(D|H_2,\hat{p}_2)}.
\end{equation}
Here we are using a set $D$ of $N$ data to compare hypotheses $H_1$ and $H_2$, and $\hat{p}_{1}$, $\hat{p}_{2}$ are the maximum likelihood estimators of their parameters. Let us call $\nu_i$ to the number of parameters under hypothesis $i$. By convention, we will reserve the denominator for the hypothesis with the largest number of parameters, i.e.\  $\nu_2 \geq \nu_1$. Hence, in Eq.\ (\ref{lambda}) the truncated power law appears in the numerator and the truncated lognormal in the denominator, while, in Eq.\ (\ref{lambda'}), the truncated lognormal is in the numerator and the double power law is in the denominator.

The likelihood ratio is meaningful for the two main schools of statistics, the Bayesian and the frequentist one. 

Bayesian statistics \cite{Jaynes2003} assigns probabilities to different hypotheses using the equations that follow mathematically from the basic axioms of probability, but its use is not trivial because of its reliance on \textit{prior probabilities}. In principle, when analysing some given data with Bayesian tools, the starting point is a set of prior probabilities, which quantify the plausibility of each hypothesis based on all the available information except these data. The probabilities that result from combining the prior probabilities with the data are named posterior probabilities. On their turn, these posterior probabilities can be used as prior probabilities when analysing new data. However, prior probabilities are often difficult to assign and there is no consensus on the best way to do it \cite{Kass1996}.

Frequentist statistics, which is the mainstream school, comprises a set of recipes that avoid using prior probabilities. In exchange, it does not have the logical transparency of Bayesian statistics, and it does not assign probabilities to hypotheses. Instead, its results are expressed using concepts such as significance levels and (in the case of parameter values) confidence intervals, which have a less clearcut interpretation and are test-dependent.

The starting point of Bayesian statistics is the Bayes theorem
\[
P(x)P(y|x) = P(y)P(x|y).
\]
If we replace the variables $x$, $y$ by a data set $D$ and a hypothesis $H$, we have
\[
P(H|D) \propto P(H)P(D|H),
\]
where $P(H)$ is the prior probability, $P(H|D)$ is the posterior probability, and $P(D|H)$ is the likelihood function. When comparing the probabilities of two different hypotheses:
\begin{equation}
\label{probratio}
\frac{P(H_1|D)}{P(H_2|D)} = \frac{P(H_1)}{P(H_2)}B,
\end{equation}
where $P(H_1)/P(H_2)$ is a ratio of prior probabilities and $B$ is the Bayes factor \cite{Kass1995,Goodman1999b}, i.e.\  the ratio of likelihood functions:
\begin{equation}
\label{Bayesfact}
B = \frac{f(D|H_1)}{f(D|H_2)}.
\end{equation}
The Bayes factor contains all the evidence extracted from the data and can be quantified without need of assigning prior probabilities to the two hypotheses. However, we still need some other prior probabilities: note that the difference between the likelihood ratio $\lambda$ in Eq.\ (\ref{lambdagen}) and the Bayes factor $B$ in Eq.\ (\ref{Bayesfact}) is that the later does not include the maximum likelihood estimates of the parameters. This is because there is no reason a priori (before analysing the data) for the parameters to take precisely these values. All possible values of the parameters have to be considered a priori \cite{Kass1995}:
\[
B=\frac{ \int f(D|H_1,p_1)f(p_1)dp_1}{ \int f(D|H_2,p_2)f(p_2)dp_2},
\]
where the prior probability distributions of the parameters appear as an input. Note that, if a hypothesis has many parameters, it is less likely a priori that all of the parameters have values that fit the data well, and, as a result, the presence of unneeded parameters decreases the probability of the hypothesis. This is not evident from the likelihood ratio in Eq.\ (\ref{lambdagen}), which, if taken at face value, favours hypotheses with many parameters.

Unfortunately, there is no way to calculate $B$ exactly without introducing prior probability distributions for the parameters. However, if we have enough with an approximate value of $B$, and the number of data $N$ is large, it is generally possible to avoid considering these distributions explicitly. Schwarz \cite{Schwarz} showed that, in broad conditions, $\ln(B)$ is roughly approximated by
\[
S = \ln(\lambda) - \frac{1}{2} (\nu_1-\nu_2) \ln(N).
\]
In the case of the truncated power law and the truncated lognormal, $\nu_p-\nu_l=-1$ and, therefore, $S = \ln(\lambda) - \ln(N)/2$. If we replace $B$ by $\exp(S)$ in Eq.\ (\ref{probratio}) we obtain a rough approximation to the relative probability of the two hypotheses. Note that if, for example, both are considered equally likely a priori, $B$ equals the ratio of probabilities and $\exp(S)$ approaches it in order of magnitude. 

In the last case (equal prior probabilities), or whenever the data have more weight than the prior probabilities in Eq.\ (\ref{probratio}), the hypothesis of choice is the one that maximizes $S$, or, equivalently, the one that minimizes
\[
BIC = -2 \ln(f(D|H,\hat{p}) + \nu \ln(N),
\]
where $BIC$ stands for \textit{Bayesian Information Criterion} or \textit{Information Criterion B}, and is expressed in this way in analogy to Akaike's \cite{Akaike} \textit{Information Criterion A} ($AIC$), widely used to compare hypotheses. Among several competing hypotheses, Akaike's criterion selects the one that minimizes
\[
AIC = -2 \ln(f(D|H,\hat{p})+2\nu.
\]
According to Schwarz's results, Akaike's criterion will tend to overestimate the number of parameters \cite{Kass1995}. $AIC$ is not used in this paper.

The likelihood ratio is also used by the frequentist school (although, in this case, it is just one of several possible statistics) when comparing nested hypotheses, such as the truncated power law and the truncated lognormal. Wilks \cite{Wilks} showed that, if the nested hypothesis with less parameters is correct and $N$ is large, $-2\ln(\lambda)$ follows a chi-square distribution with $\nu_2-\nu_1$ degrees of freedom. For example, when comparing the truncated power law to the truncated lognormal, the power law can be rejected with a significance level 0.05 when $\lambda < 0.147$, or with a significance level 0.01 when $\lambda < 0.036$.

Both Schwarz \cite{Schwarz} Bayesian criterion and Wilks \cite{Wilks} frequentist criterion are taken into account in this study.

\section{Calculation}

\subsection{Data}
\label{Data}

The data on economic contractions were obtained from the Barro-Urs\'{u}a Macroeconomic Dataset \cite{DadesUrsua}, which comprises time series of the annual per capita GDP from 42 countries. These series cover from the earliest reliable data available in each country (most often in the 19th century) to 2009. Following ref.\ \cite{BarroUrsua}, three countries were excluded because of missing data around World War II: Malaysia, Philippines and Singapore. 

For the remaining 39 countries, the relative top-to-trough contraction size was obtained for each of 691 country-level downturns. For the purposes of this study, the top-to-trough distance $\Delta GDP$ was calculated for events spanning since per capita GDP begins to decrease until it recovers its former level (without considering nested events), and divided by the initial its initial figure, i.e.\ $x=-\Delta GDP/GDP$ (referring, as throughout the paper, to per capita GDP).

This type of measure is useful because of its simplicity, transparency and tractability, in spite of the important drawback that, whenever a downturn affects several countries, it is treated as a set of independent contractions. A related problem is that the statistical treatments in this paper assume independence, but these data are not fully independent. However, given the large number of data and the diversity of situations (many OECD and non-OECD countries, different historical periods), I assume that the covering of the sampling space is broad enough to compare to a fully random sample, at least when identifying the type of distribution, which is probably more robust than the specific values of the parameters.

At different stages of the application of TASEPOLAR, either the whole dataset or subsets are sorted in either ascending or descending order. However, as a way to identify particular data unambiguously, they are always labelled according to their position when the whole dataset is sorted in descending order, i.e.\ the largest contraction is \#1 and the smallest is \#691.

\subsection{Application of TASEPOLAR to economic contractions}
\label{aplic}

\subsubsection{Identification of the power law range}
\label{ident}

First, I selected, as a seed, a range of values that is well fitted by a power law. This choice is rather arbitrary. The range should be relatively large but it should leave, on both sides, sequences of data that are also likely to make part of the power law. I selected it tentatively by visual inspection of the empirical PDF, and then I tested its goodness of fit. The chosen range was $D_{80,197}$. In this range, the likelihood ratio between the truncated power law and the truncated lognormal is 0.79. This is well within the range of values that we expect if the power law hypothesis is correct: according to Wilks' approximation, under this hypothesis there is a probability 0.49 of $\lambda_{80,197} \leq 0.79$.

\begin{figure}[t]
\begin{center}
\includegraphics[scale=0.7]{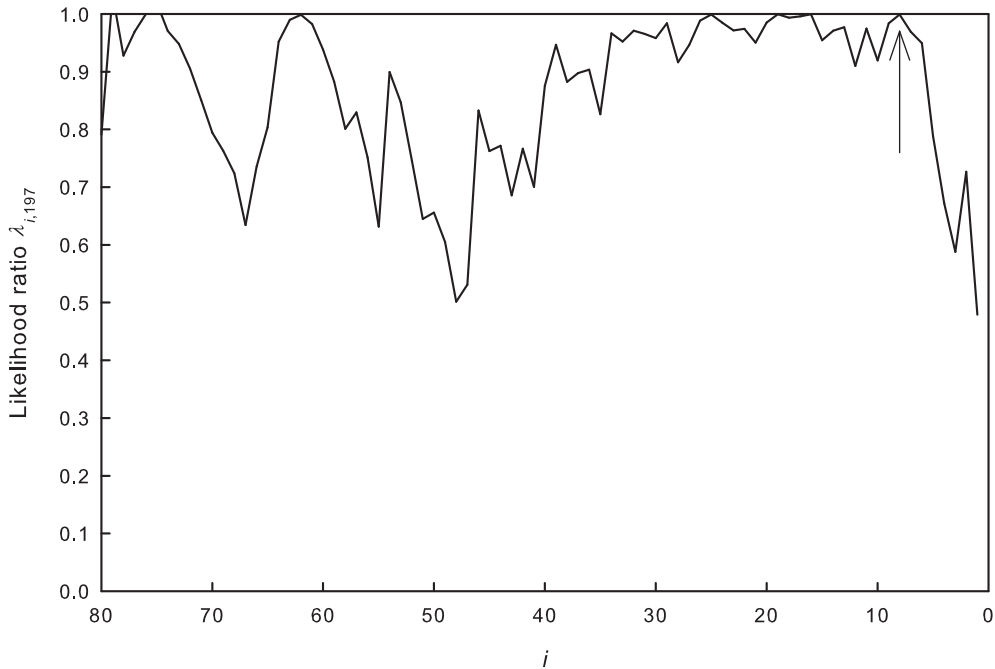}
\end{center}
\caption{First pointwise sequential comparison between the power law and the truncated lognormal. Likelihood ratio $\lambda_{i,197}$ between the power law and the truncated lognormal from $i=80$ to $i=1$, i.e.\ for the sequence of nested data sets from $D_{80,197}$ to $D_{1,197}$. Since these are nested distributions, $\lambda$ cannot take any value larger than 1. The arrow indicates the point chosen as the upper limit of the power law. Even though the ensuing decline is not larger than we could expect from a random fluctuation in a power law, in this case it probably denotes a real deviation from this distribution, because the power law necessarily has an upper limit.}
\label{figseq1}
\end{figure}

Second, I performed a pointwise sequential comparison to expand the selected range on the upper side. Figure \ref{figseq1} gives the likelihood ratio for each of the nested subsets, from $D_{80,197}$ to $D_{1,197}$. Most of the sequence displays minor fluctuations below $\lambda = 1$, as we expect from a power law. The eighth datum before the end ($x_8 = 0.5034$) is the last whose inclusion gives $\lambda > 0.99$. In fact, it gives a value almost equal to one: $\lambda_{8,197} = 0.99906$. The sequential inclusion of the 7 remaining values gives what appears to be a downward trend culminating at $\lambda_{1,197} = 0.4790$. This $\lambda$ is not extremely low and could have resulted from a random fluctuation (the frequentist Wilks' criterion does not allow rejecting the power law for any significance level $< 0.23$). For $\lambda_{1,197}$, Schwarz statistic is $S = -1.9$, giving $\exp(-1.9)=0.15$ as an approximation to the Bayes factor, in favour of the power law. However, the Bayes factor is to be combined with the prior probabilities (Eq.\ \ref{probratio}), and we know that the power law has to bend at some point because it cannot continue beyond the total size of the system, so there is a high prior probability that a few of the largest observed values do not make part of the power law. Therefore, we do not assume that the values $x_1$ - $x_7$ are included in the power law range. The estimated $x_{max}$ is thus $x_8=0.5034$. Since the previous value in the sequence is $x_9 = 0.4987$ and the next is $x_7=0.52714$, $x_{max}$ is not significantly different from 0.5.

\begin{figure}[!b]
\begin{center}
\includegraphics[scale=0.7]{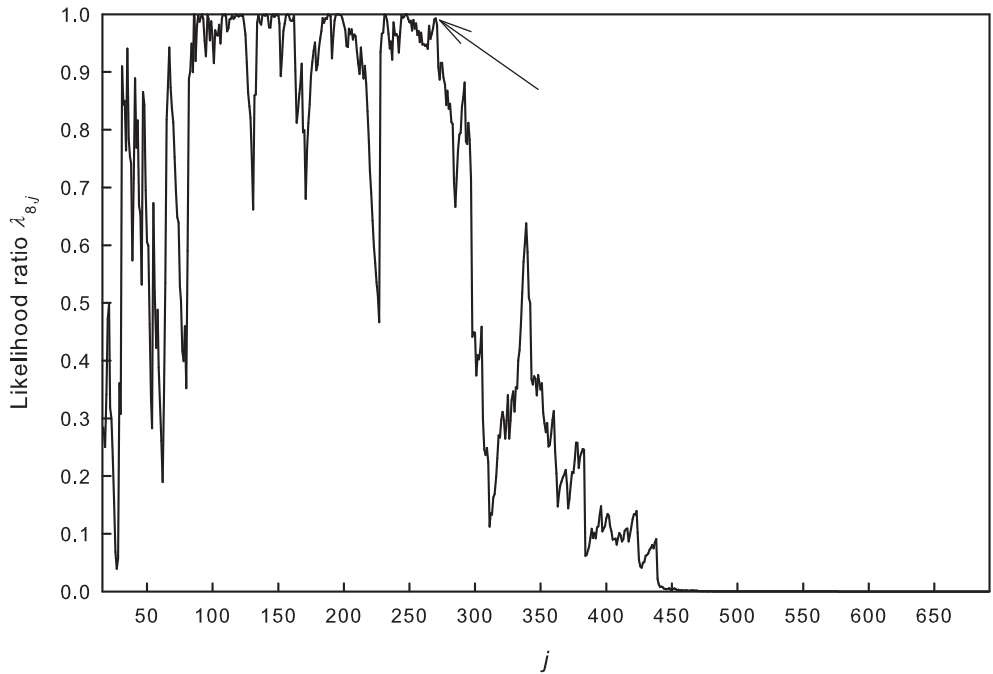}
\end{center}
\caption{Second pointwise sequential comparison between the power law and the truncated lognormal. Likelihood ratio $\lambda_{8,j}$ between the power law and the truncated lognormal from $j=17$ to $j=691$, i.e.\ for the sequence of nested data sets from $D_{8,17}$ to $D_{8,691}$. Since these are nested distributions, $\lambda$ cannot take any value larger than 1. The arrow indicates the point chosen as the lower limit to the power law.}
\label{figseq2}
\end{figure}

Third, I performed the definitive pointwise sequential comparison. After having estimated $x_{max}$, I took the 20 largest values satisfying $x \leq x_{max}$, i.e.\ $D_{8,27}$, and I enlarged this subset progressively. Figure \ref{figseq2} gives the likelihood ratios for all the nested subsets from $D_{8,27}$ to $D_{8,691}$ (recall that $x_{691}$ is the smallest contraction in the sample). After some large fluctuations which were expected because of the small sample size, we find again a region with small, steady fluctuations slightly below $\lambda = 1$. The range that I first chose as a seed is within this region, implying that it was a valid choice. At some point, a decreasing trend appears. When the sequence is complete, we are left with $\lambda_{8,691} = 7.5 \times 10^{-53}$. In frequentist terms, this allows rejecting the power law with a significance level $4 \times 10^{-54}$. Schwarz's statistic is $S = -117$, which gives $\exp(S) \approx 2 \times 10^{-51}$ as an approximation to the Bayes factor. This makes it virtually certain that this whole range cannot be fitted by one single power law. The last datum giving $\lambda > 0.99$ ($\lambda_{8,270} = 0.9931$) was selected as the lower bound to the power law. This value is $x_{270} = 0.05327$, i.e.\ $x_{min} \approx 0.053$.

Thus, TASEPOLAR gives a range for the power law covering from $x_{min} = 0.053$ to $x_{max} = 0.5$, and comprising 263 data, or 39.5\% of the sample, while allowing us to definitely discard the hypothesis of one single power law covering the whole data set. The maximum likelihood estimate of the power law exponent in the selected range is 1.77.

\subsubsection{Test of the power law hypothesis}
\label{test}

Here I test whether our candidate power law range, covering the data \#8-\#270, is something else than a slice of a truncated lognormal. As described in Section \ref{basic}, we do this by testing whether an enlarged region can be interpreted as two slices of a single truncated lognormal. In this study, we enlarge the region in one of the simplest manners, by including all the smaller values, i.e.\ we analyse $D_{8,691}$. In this region, we compare the truncated lognormal with a double power law.

Before proceeding, we want to make sure that the exclusion of the few largest data (\#1-\#7) is not unfair toward the lognormal hypothesis. In the case of the power law, an upper truncation was unavoidable because, otherwise, its mean would be infinite (for the estimated exponent), i.e.\ infinitely above the maximum possible value $x = 1$ (corresponding to complete destruction). The lognormal decays quicker than the power law, hence it is possible a priori that its probability density becomes negligible before reaching $x = 1$, in which case an explicit truncation would not be needed. However, for the fitted parameters in our case (see below), if there were no truncation there would be a probability 0.025 that any given contraction were larger than one. As a consequence, there would be a probability of only $2.5 \times 10^{-8}$ that none of our 691 data crossed this threshold, which means that an explicit truncation is needed also for the lognormal.

Maximum likelihood estimation of the parameters of the truncated lognormal in the enlarged range gives $\hat{\tau} = 1.17$ and $\hat{\psi} = 1.18$ (equivalent to $\hat{\mu} = -3.29$, $\hat{\sigma} = 1.68$). By applying maximum likelihood estimation to the double power law, the boundary between the two power laws appears located at $x_{333} = 0.037$. The estimated exponent above this value is $\hat{\tau}=1.63$ (smaller than estimated in Section \ref{ident}, because of the inclusion of smaller values, which might actually deviate from the power law), while, below, it is $\hat{\tau}'=0.31$.

The likelihood ratio between the truncated lognormal and the double power law (Eq.\ \ref{lambda'}) is $2.2 \times 10^{-6}$, favouring the double power law. Because these are not nested models, this number cannot be translated to a frequentist significance level using Wilks criterion. However, it can be given a Bayesian interpretation. Schwarz's statistic is $S = - 9.76$, with $\exp(S) = 5.8 \times 10^{-5}$ being an approximation to the Bayes factor $B$ (Eq.\ \ref{Bayesfact}). Even though there is much error in the translation from $S$ to $B$, the magnitude of this result makes it virtually certain that the double power law fits the data better than the truncated lognormal.

The truncated lognormal could only beat the double power law if it were assigned a prior probability several orders of magnitude above the double power law, but there is no reason for this, because there are plausible mechanisms that can produce a power law in our context, as discussed in Section \ref{discusmec}. Therefore, if we are to decide whether our data follow a truncated lognormal or a double power law distribution, the second option is more probable by several orders of magnitude. This does not necessarily mean that the double power law gives an optimum description of the data, because we are just comparing two distributions that were chosen, in part, because of their simplicity, while the set $D_{8,691}$ could have a more complex shape. However, this result lends much support to the hypothesis that the power law is the optimum function to fit the region selected in Section \ref{ident}, comprising from $x_{min} \approx 0.053$ to $x_{max} \approx 0.5$.

\subsection{Complementary analyses}

\subsubsection{Plotting the PDF}
\label{plot}

The empirical probability density function was plotted with logarithmic binning. Reference \cite[pp. 131-132]{Pueyo2003}, describes this method (already used as early as ref.\ \cite{Manna1990}) and shows that it is optimum for power laws and related distributions. In order to give a clear image of the power law, the range between the estimated $\log(x_{min})$ and $\log(x_{max})$ was divided into seven equal bins. The binning was also extended to smaller (in Fig.\ \ref{figtot}) and larger (in Fig.\ \ref{figpdf} and Fig.\ \ref{figtot}) values.

Along with the empirical PDF, Fig.\ \ref{figpdf} displays the fitted power law as a line crossed by several segments representing probability intervals. The data points of the empirical PDF are expected to lie within these intervals with probability $2/3$. The apportionment of data among bins should follow a multinomial distribution, which I approximate as one binomial distribution for each bin. If a bin runs from $x_A$ to $x_B$, the probability for a contraction to make part of this bin is $F(x_B)-F(x_A)$, where $F$ is the cumulative probability function of the power law. This probability is the parameter used for the binomial. Usually, no range of $x$ will give exactly a probability of $2/3$, because the distribution of the number of events is discrete. However, I selected the values of $x$ giving the immediately smaller and larger probabilities, and I obtained the intervals from them by linear interpolation.

\subsubsection{Power law exponent}
\label{secexpon}

In a frequentist approach, the exponent $\tau$ of the power law (Eq.\ \ref{powerlaw}) is obtained by maximum likelihood estimation and can be complemented with confidence intervals. It is straightforward to show that the maximum likelihood estimator of $\tau$ is the value $\hat{\tau}$ that satisfies:
\begin{equation}
\label{mle}
\overline{\ln(x)}=(\hat{\tau}-1)^{-1}+\frac{x_{min}^{-\hat{\tau}+1}\ln(x_{min})-x_{max}^{-\hat{\tau}+1}\ln(x_{max})}{x_{min}^{-\hat{\tau}+1}-x_{max}^{-\hat{\tau}+1}}.
\end{equation}
From a Bayesian point of view, if we are to estimate $\tau$ from a data set $D$ we have to assign a prior probability distribution $f(\tau)$:
\[
f(\tau|D) \propto f(\tau)f(D|\tau).
\]
Based on the fact that most empirical data on other types of catastrophic events and most models attempting to predict them involve values of $\tau$ somewhere between 1 and 2 or close to this range \cite{Pruessner}, an expert prior should probably be a broad distribution with its mode between 1 and 2. Since (i) the derivative is zero at the mode, (ii) the distribution should be broad, and (iii) the precise location of the mode is unclear, a uniform distribution is well-justified for values that do not deviate strongly from the range 1-2. This requirement is not problematic in our case, because the likelihood function that we find is narrow and centered in a value within this range. This simplifies our calculations because, for a uniform prior, the posterior probability $f(\tau|D)$ is proportional to the likelihood function $f(D|\tau)$. Furthermore, since confidence intervals are also obtained from the likelihood function, the difference between Bayesian cumulative probabilities and frequentist confidence intervals is minimum in this case.

More problematic is the fact that the data are not fully independent. However, as the characterization of their interrelation is far from trivial, I only obtained provisional results under the assumption of independence. Since $\overline{\ln(x)}$ in Eq.\ (\ref{mle}) is the average of a large number of variables with a definite mean and variance, it will be approximately Gaussian. For each possible value of $\tau$, $\ln(x)$ will follow a Gaussian distribution of mean $\overline{\ln(x)}$ in agreement with Eq.\ (\ref{mle}), and variance $[\overline{\ln(x)^2}-\overline{\ln(x)}^2]/\sqrt{N}$, where
\begin{eqnarray*}
\overline{\ln(x)^2} = & ([\ln(x_{min})^2+\frac{2}{\tau-1} \ln(x_{min})+\frac{2}{(\tau-1)^2}]x_{min}^{-\tau+1}-[\ln(x_{max})^2 \\
& +\frac{2}{\tau-1} \ln(x_{max})+\frac{2}{(\tau-1)^2}]x_{max}^{-\tau+1})/(x_{min}^{-\tau+1}-x_{max}^{-\tau+1}).
\end{eqnarray*}
The Gaussian density function thus calculated for each $\tau$ is the likelihood function of that $\tau$.

\subsubsection{Comparison between the power law and the generalized Pareto distribution type III}
\label{rescomp}

\begin{figure}[!b]
\begin{center}
\includegraphics[scale=0.7]{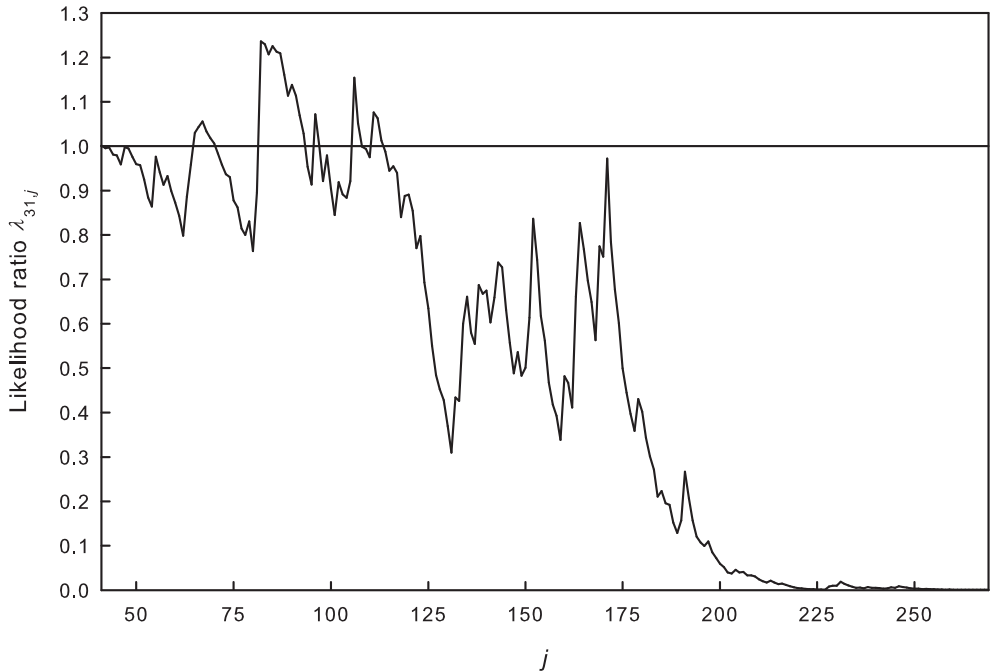}
\end{center}
\caption{Pointwise sequential comparison between the generalized Pareto distribution type III (gPd3) and the power law. Likelihood ratio $\lambda_{31,j}$ between the gPd3 and the power law from $j = 40$ to $j = 270$. Since these are not nested distributions, $\lambda$ can take values larger than 1.}
\label{figgpd}
\end{figure}

An earlier version of the dataset used in this study was analysed by Barro and Jin \cite{BarroJin}. These authors attempted to fit a power law to $y = 1 / (1-x)$. Since $f(x) = f(y)|dy/dx|$, the power law $f(y) = a y^{-\tau}$ corresponds to the following distribution in terms of $x$:
\[
f(x) \propto (1-x)^{-\gamma},
\]
which is a generalized Pareto distribution type III (gPd3) with exponent $\gamma = \tau - 2$ (the power law is a generalized Pareto distribution type II, which corresponds to the original Pareto distribution) \cite{Pickands}.

Barro and Jin \cite{BarroJin} tested this distribution by visual inspection of the cumulative probability plot of $y$. In view of a notable disagreement between the expected and obtained plot, they split the set of data $x$ that they were analysing (all of them satisfying $x > 0.1$, which was a criterion set a priori by the authors) into two ranges: (1) $0.1 < x < 0.32$, and (2) $x > 0.32$. I compare the power law with the gPd3 in several ranges, considering both Barro and Jin's criteria and my estimates above. In all cases, I consider truncated versions of both distributions.

For $D_{1,7}$ (i.e.\ for $x>0.5$, corresponding to the large values that we excluded from the power law), the likelihood ratio between the gPd type III and the power law is 0.95. Schwarz's criterion would give a Bayes factor similar to this ratio if it could be applied, because the number of parameters is the same for both hypotheses, but it cannot be applied because the number of data is too small. Therefore, we cannot discriminate between the two hypotheses based on the data. However, the gPd type III is more likely a priori because we expect the upper range discarded by TASEPOLAR to correspond to the part of the distribution that does not follow a power law because of the proximity to the upper bound ($x = 1$), and the gPd type III is precisely a distribution used for bounded variables. Wilks' frequentist criterion cannot be applied, not only because of the few data but also because these are not nested distributions.

For $D_{1,30}$ (corresponding to the range $x > 0.32$ selected by Barro and Jin, and comprising both data included and excluded from the power law range by TASEPOLAR), the likelihood ratio is 0.51. This suggests some advantage for the power law (in a relation 2:1 in favour of this distribution), but, again, the number of data is too small.

The second range considered by Barro and Jin \cite{BarroJin} was $x < 0.32$. I performed a pointwise sequential comparison contemplating this upper bound with the lower bound that they chose by convention ($x = 0.1$) but also larger and smaller lower bounds, down to the smallest value that TASEPOLAR includes in the power law ($x = 0.053$). This pointwise comparison, spanning from $D_{31,40}$ to $D_{31,270}$, is shown in Fig.\ \ref{figgpd}. The figure indicates that the two distributions cannot be distinguished when taking few data, but a larger sample (associated to a broader range) gives advantage to the power law. For $D_{31,270}$, the likelihood ratio between the gPd type III and the power law is $7.8 \times 10^{-4}$, which allows us to neatly reject the gPd type III in favour of the power law.

\section{Results}
\label{results}

\begin{figure}[t]
\begin{center}
\includegraphics[scale=0.7]{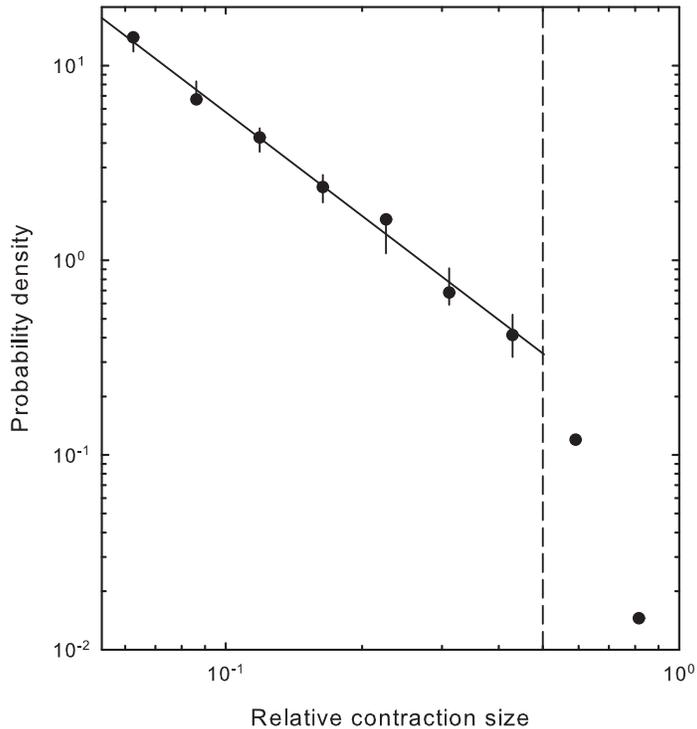}
\end{center}
\caption{GDP contraction size distribution. Probability density function (PDF), on a log-log scale, of the relative sizes of 270 country-level per capita GDP contractions representing at least a 5.53\% loss. A power law gives an optimum fit from 5.53 to 0.5\% loss (vertical dashed line). The dots correspond to the empirical PDF, represented with logarithmic binning. The continuous line is the best-fit power law. The error bars are expected to embrace the empirical observations with probability $2/3$ (Section \ref{plot}).}
\label{figpdf}
\end{figure}

\begin{figure}[t]
\begin{center}
\includegraphics[scale=0.55]{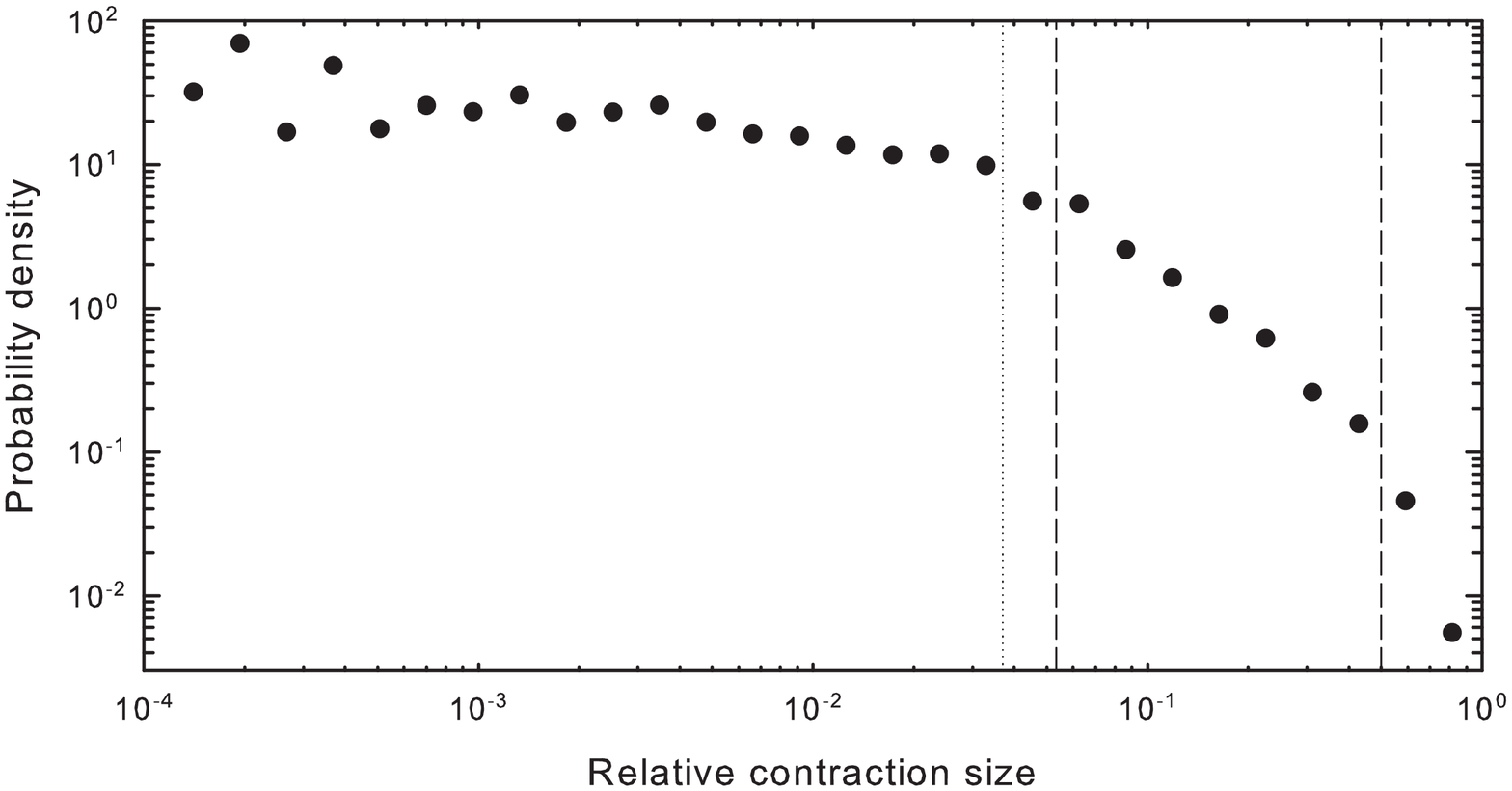}
\end{center}
\caption{Enlarged GDP contraction size distribution. Empirical PDF, on a log-log scale, of the relative sizes of the 691 country-level per capita GDP contractions in the sample, using logarithmic binning. The vertical dashed lines indicate the part of the distribution that TASEPOLAR identifies as a power law (Section \ref{ident}), which sets the lower limit for Fig.\ \ref{figpdf}. The dotted line indicates the less conservative lower limit that results from the criterion applied in Section \ref{test}. The range below the power law includes values as close to zero (i.e.\ as negative on a logarithmic scale) as allowed by the resolution of the data, which determines the length of this range.}
\label{figtot}
\end{figure}

TASEPOLAR detected a well-defined power law in the contractions causing from 5.53 to 50\% per capita GDP loss. This range includes 263 events, i.e.\ 38\% of the 691 contractions in the sample (61\% are smaller, 1\% are larger). Figure \ref{figpdf} shows the fitted and the empirical PDF for losses $\geq 5.53\%$. To contextualize Fig.\ \ref{figpdf}, the complete distribution is given in Fig.\ \ref{figtot}.

The best-fit exponent $\hat{\tau}$ is 1.77. Ignoring correlations, this value is estimated to lie in the range (1.61,1.94) with probability 0.9 (this is a Bayesian probability interval; in frequentist terms, the 90\% confidence interval is (1.60,1.94)), but correlations might make this range larger.

In addition to applying the tests that constitute the backbone of TASEPOLAR (Section \ref{aplic}), the distribution used in ref.\ \cite{BarroJin} was also tested and was clearly rejected in front of the power law (apart from the few data at the upper end, where the data do not allow discriminating between both distributions but a power law is unlikely).

\section{Discussion}

\subsection{Is it a power law?}

TASEPOLAR (in its first part, Section \ref{ident}) identified a range of values of relative per capita GDP contractions that is well-fitted by a power law. The way TASEPOLAR is designed nearly ensures that no other function can fit the data better in this range. The bend downwards in the upper end of Fig.\ \ref{figpdf} is unavoidable as we approach size 1, corresponding to the complete destruction of the system. As mentioned in Section \ref{basic}, a deviation from the power law at small values is also unavoidable, because, otherwise, the size of the contractions would be 0 with probability 1.

Possible replies are that the power law range covers only one order of magnitude (from 5.53 to 50\%) and that this range appears small in the context of the complete distribution in Fig.\ \ref{figtot}. However, there are several reasons to consider this power law meaningful: (1) In spite of the relatively short range, the tight probability intervals in Fig.\ \ref{figpdf} leave little margin for deviations from a straight line. (2) There is no evident way to subsume the selected range in a simple function embracing the complete data set in Fig.\ \ref{figtot}; the alternative to the power law that was considered in the second part of TASEPOLAR (Section \ref{test}) was outright rejected. (3) Even though the power law range appears relatively small in the context of the complete distribution in Fig.\ \ref{figtot}, this is misleading because the range below the power law includes values as close to zero (i.e.\ as negative on a logarithmic scale) as allowed by the resolution of the data, which determines its length; in terms of number of events, a relatively large share (38\%) belong to the power law range. (4) The power law range is economically important because it embraces nearly all depressions (except a few extreme events on the upper end) if, as it is often done, we use a 10\% threshold to define \textit{depression}. This threshold implies that, by definition, the distribution of depression sizes cannot exceed one order of magnitude; however, our results point to the power law as the optimum function to describe them, and also to describe smaller contractions up to half this minimum size.

\subsection{Data treatment methodology for power laws}
\label{discusmet}

In this study I developed a method to identify and fit the power law distribution, and applied it to relative GDP contractions.

The few earlier attempts to fit a power law to GDP contractions used simpler methods. Most often, power laws are identified from visual inspection of log-log plots of various types, but this has severe limitations \cite{Clauset}. Ormerod and Mounfield \cite{Ormerod2001} were probably the first to investigate the hypothesis of a power law in the sizes of economic contractions (measured in a way slightly different from ours), but their results were inconclusive. These authors binned the data linearly, and explored the adequacy of the power law using regression and visual inspection on a plot of expected vs found values. Observing a poor agreement, they removed a few of the largest data and noted that the plot somewhat improved. However, linearly binning a power law causes a large bias, both in regressions and in visual inspection \cite{PueyoJovani2006}. It is especially difficult to extract information from a plot like the one obtained by these authors, which is dominated by a few dots while the rest form a cloud. In addition, the full set of contractions was used, which amounts to setting zero as the lower bound. A power law (with $\tau \geq 1$) cannot extend to zero (Section \ref{basic}); whether or not a plot extending to zero reminds a power law depends strongly on binning details (as shown in ref.\ \cite{PueyoFearnside2011} in a different context). In spite of these limitations, which were difficult to overcome given the state of the art just a few years ago, Ormerod and Mounfield's pioneering analysis was essential as an initial exploration and in pointing the way to further research. Similar comments apply to ref.\ \cite{Ormerod2004}. In ref.\ \cite{Xi} an empirical histogram of contraction sizes was also shown to display some rough similarity to a power law, but no other test was carried out.

More complex methods to deal with power laws have been developed in different contexts. Several of these, like TASEPOLAR, test the power law for several possible $[x_{min},x_{max}]$ in order to choose one. However, most previous methods were devised to find only $x_{min}$, usually, assuming $x_{max} \rightarrow \infty$, which, as shown above, is unrealistic.

TASEPOLAR builds on the sequential pointwise goodness-of-fit test and model comparison method that the author developed previously in a different context (summarized in ref.\ \cite[p. 136]{Pueyo2003}). In that work, for each $j$, the goodness of fit to the power law and to a different distribution (the negative exponential) were tested for $D_{1,j}$, using Snedekor's test, and the results were plotted as a function of $j$. The plot allowed identifying the ranges in which each of the hypothesis could not be rejected, and the hypothesis covering a broader range was selected. With the help of the author, Bartumeus et al.\ \cite{Bartumeus} also applied a sequential pointwise model comparison, using Akaike's information criterion \cite{Akaike} ($AIC$) with corrections, and, separately, a sequential goodness-of-fit test, which, following Clauset et al.\ \cite{Clauset}, was based on the Kolmogorov-Smirnov statistic $KS$. 

On their part, Clauset et al.\ \cite{Clauset} chose as $x_{min}$ the $x_{j}$ minimizing the $KS$ statistic obtained from $D_{1,j}$ under the assumption of a power law. In addition, they tested the goodness of fit by comparing the minimum $KS$ obtained from the empirical data with the minimum $KS$ obtained from each of a set of synthetic data sets. These data sets were produced based on the hypothesis of a power law tail with the best-fit $\tau$ and $x_{min}$, while values $x<x_{min}$ were generated by bootstrapping. Clauset et al.'s \cite{Clauset} is a frequentist method related to Handcock and Jones \cite{Handcock} Bayesian method. The later authors used the Bayesian Information Criterion ($BIC$, see Section \ref{interpret}) to compare and test a set of hypotheses corresponding to different $x_{min}$, assuming a power law for $x \geq x_{min}$ and an additional parameter for each of the values $x<x_{min}$ found in the original sample. As noted by Clauset et al., though, this method overestimates the size of the power law range because it overestimates the number of parameters needed to describe the other part of the distribution, since the $BIC$ criterion penalizes unneeded parameters. Having noted some problems with Clauset et al.'s method, Deluca and Corral \cite{DelucaCorral2013} introduced some refinements; furthermore, they extended it to estimate also $x_{max}$.

Therefore, the main features that distinguish TASEPOLAR are: (1) $x_{max}$ is estimated in addition to $x_{min}$, as is needed for fundamental reasons (Section \ref{basic}; however, this feature is shared by ref.\ \cite{DelucaCorral2013}); (2) it solves the problem of the hypothesis with which the power law should be compared by choosing a hypothesis that implies the least possible assumptions, based on a Taylor series expansion; (3) even though it also accommodates frequentist criteria, it has a Bayesian backbone that allows assigning probabilities, rather than relying on frequentist tests, which demand arbtrary choices and do not assign probabilities; (4) it avoids built-in biases, such as those implicit in $AIC$ (which, apparently, overestimates the number of parameters; Section \ref{interpret}) or those that arise when applying $BIC$ under the hypothesis choice in ref.\ \cite{Handcock}.

However, TASEPOLAR still has room for improvement. First, it gives no systematic recipe for the distribution that should be assumed for $x \leq x_{min}$ under the hypothesis of a power law for $x_{min} \leq x \leq x_{max}$ (neither it does for $x \geq x_{max}$, but this appears to be less critical). Second, it approximates the Bayes factor by means of the likelihood ratio, but the Bayes factor could also be calculated directly after a sound choice of prior probabilities for the parameters.

\subsection{Interpretation and consequences of the power law}
\label{discusmec}

Models for power-law distributed catastrophic events abound in the complex systems literature. Because of some shared features, the dynamics of many of these models have been interpreted as different instances of a single phenomenon known as \textit{self-organized criticality} (SOC) \cite{Pruessner}. Power laws in catastrophic events are often attributed to SOC, even though this is not always correct \cite{Pruessner,Pueyoetal2010}. SOC economic models in the literature \cite{Bak_econ,Scheinkman,Krugman,Thurner,Xi} certainly anticipated the results found in this paper, but we do not know whether they did it for the right reasons. These are simple models, but are thought to be potentially useful because the outcome of SOC models is nearly independent of their details if some basic requirements are met \cite{Pruessner}.

One basic ingredient of SOC is the propagation of well-defined disturbances from unit to unit in the form of chain reactions \cite{Pruessner}. Some classical studies interpreted economic downturns as chains of bankruptcies propagating from firm to firm in various ways \cite{Marx}, and such chains have met renewed interest \cite{Battison,DelliGatti2009}. However, we do not know how important this \textit{microscopic} transmission is as compared to effects that take place directly at the level of macroeconomic variables. Furthermore, an attempt to identify chains of bankruptcies \cite{Fujiwara2010} resulted into empirical PDFs that do not appear to correspond to power laws, although this could be due to the fact that not all modes of transmission could be tracked, and that the study only comprised one year, while chains can last much longer \cite{Fujiwara2012}. 

Another basic ingredient is a negative feedback. Contagion should become more difficult immediately after a chain reaction, but, later, firms should progressively recover their vulnerability. This is plausible, because, in periods of stability, firms could diversify into multiple ways of functioning, some of which would not be viable in periods of turmoil. These firms would have become vulnerable for a variety of reasons, such as high specialization, high leverage (Minsky's \cite{Minsky} mechanism), or stretched budgets (inter alia, because of increasingly obsolete products or means of production, setting the basis for \textit{gales of creative destruction} \cite{Schumpeter}).

One third ingredient is time scale separation. Chain reactions should be brief as compared to the time needed for firms to become vulnerable again. In particular, there should be little chance of a chain reaction affecting companies born (or recovering from bankruptcy) once the reaction had already started. The observation of chain reactions extending over several years \cite{Fujiwara2012} means that this cannot be taken for granted.

Only a deeper investigation of each of the above items will allow determining whether the power law in Fig.\ \ref{figpdf} results from SOC. Besides firms, products could also be the units of SOC chain reactions \cite{Thurner,Xi}: this hypothesis has empirical support \cite{Klimek}. SOC or not, it is clear that fluctuations propagate through channels not only internal but also external to the market. Notably, some of the largest contractions in the dataset are related to the two world wars and other conflicts \cite{BarroUrsua}.

The granular theory of economic fluctuations \cite{Gabaix_granular} is also consistent with the results in the present paper. This approach suggests that scale-invariant fluctuations occur as a consequence of scale-invariance in the sizes of pieces making up the economy, such as firms or individual incomes. Each of these pieces is susceptible to suffering shocks, whose importance for the whole economy depends, in part, on the size of the piece. SOC can be considered a particular instance of this mechanism, one particularity being that, in SOC systems, the scale-invariant structure is not only a cause but also a result from scale-invariant fluctuations, i.e.\ there is a feedback between both aspects of the system \cite{Pueyo2007}. Apart from SOC, there are other mechanisms that can introduce scale-invariance into the structure of the economy \cite{Dosi,Fujiwara2004,Yakovenko}, which can then translate to scale-invariant fluctuations as proposed by this theory. Refs.\ \cite{Amaral,Ormerod2004} suggest specific models that can also be considered instances of the granular theory.

Other possible explanations have a less dynamic, more statistical nature. In some cases, mixtures of distributions result into power laws \cite{Allen}. Thus, at the current stage, we cannot discard that the power law in Fig.\ \ref{figpdf} be related to the fact of mixing different countries and to the nonstationarity of their time series. However, the conditions for a mixture to give rise to a power law are not well known (a broad family of cases has been studied by the author elsewhere, but one leading to $\tau \approx 1$).

Whatever the cause, the results in this paper add unplanned economic contractions to the long list of types of \textit{catastrophic} events in which a power law distribution has been found (the analysed events can be considered \textit{catastrophic} because they were presumedly unplanned, being thus fundamentally different from planned contractions \cite{Schneider}). However, it would not be justified to leap from this observation to the conclusion that such events are so intimately related to the very nature of complex systems that there is no way to alter their distribution. Previous studies with forest fires give insights on this point. Both in SOC fire models and in empirical observations it is found that the exponent $\tau$ and the maximum fire size change predictably as a function of weather \cite{Pueyo2007}. Furthermore, both in models and in nature, there are climates in which forests adopt a radically different dynamic, with virtual absence of fire \cite{Pueyoetal2010}. Similarly, the parameters of the size distribution of economic contractions could depend on a variety of factors, ranging from natural resource availability to economic policy. The observation of a power law, which is sometimes associated to notions of universality, is not necessarily incompatible with policies that, within environmental constraints, favour a socially desirable path of change in the level of economic activity.

\bibliographystyle{unsrt}

\end{document}